\documentclass[10pt,conference]{IEEEtran}
\usepackage{algorithmic}
\usepackage[caption=false]{subfig}
\usepackage{color}
\usepackage[ruled,vlined,longend,linesnumbered]{algorithm2e}
\usepackage{graphicx}
\usepackage{epstopdf}
\usepackage{tabularx,ragged2e}
\usepackage{pdfpages}
\usepackage{multirow}
\usepackage{tabularx,ragged2e}
\newcolumntype{L}{>{\RaggedRight\arraybackslash}X}
\usepackage[numbers,sort&compress,square]{natbib}

\usepackage[hyphens]{url}
\usepackage{hyperref}
\usepackage[hyphenbreaks]{breakurl}

\begin{document}

\title{Checking App Behavior Against App Descriptions: What If There are No App Descriptions?}

    \author{\IEEEauthorblockN{Md. Shamsujjoha$^\dag$, John Grundy$^\dag$, Li Li$^\dag$, Hourieh Khalajzadeh$^\dag$, and Qinghua Lu$^\ddag$}
\IEEEauthorblockA{$^\dag$ Department of Software Systems and Cybersecurity, Faculty of Information Technology, Monash University, Australia 
\\
$^\ddag$ Data61, CSIRO,  Australia.\\
Email: \{md.shamsujjoha, john.grundy, li.li, hourieh.khalajzadeh\}@monash.edu, qinghua.lu@data61.csiro.au}
}
\maketitle

\begin{abstract}
Classifying mobile apps based on their description is beneficial for several purposes. However, many app descriptions do not reflect app functionalities, whether accidentally or on purpose. Most importantly, these app classification methods do not work if the app description is unavailable. This paper investigates a Reverse Engineering-based Approach to Classify mobile apps using The data that exists in the app, called REACT. To validate the proposed REACT method, we use a large set of Android apps (24,652 apps in total). We also show REACTs' extendibility for malware/anomaly detection and prove its reliability and scalability. However, our analysis shows some limitations in REACT procedure and implementation, especially for similar feature based app grouping. We discuss the root cause of these failures, our key lessons learned, and some future enhancement ideas. We also share our REACT tools and reproduced datasets for the app market analyst, mobile app developers and software engineering research communities for further research purposes.
\\
\end{abstract}


\section{Introduction}
\label{Sec1}
The number of smartphone users exceeded 3.8 billion in 2020, increasing by more than 40\% in the last four years. Besides, more than 205 billion mobile apps were downloaded from the cloud app repositories increasing 7.71\% per year on average since 2018~\cite{MD_SANER-2021,Confirmation_1_John,confirmation_3_Statistica,confirmation_4_ironsource}. In the last few years, Android has become the most popular platform for mobile devices due to its open-source nature, and Google Play is the official pre-installed store and portal for Android-powered devices. In the first quarter of 2018, around 3.8 million Android apps 
were available for download on Google Play Store, whereas, in the second
quarter of 2019, this number was reduced to 2.1 million~\cite{Li_And_removed}. 
Primary reasons for the removal of these apps are (i) they were flagged as malware or contain anomalies, (ii) did not behave as advertised, (iii) had low quality and designed~inappropriately, or (iv)~violated policies and were miscategorized~\cite{TSE_miscatagorization}. 

Several research efforts have been conducted to address these problems mentioned above, especially malware/anomaly detection through classification. In the CHABADA work~\cite{Chabada}, more than twenty thousand Android apps were evaluated. Initially, CHABADA extracts the topic from the app's human authored description and builds a related app cluster. Then, sensitive API usage for each cluster is grouped. Finally, if an app uses any sensitive APIs that are uncommon for its corresponding cluster, it is marked as an outlier for further analysis. A completely different approach than CHABADA to profile apps based on static information is presented in~\cite{Wei_2012}. Here, potential malicious activity is detected via changes in intensity level. A semi-supervised approach is presented in~\cite{Ma_2015} and shows that this method's performance is comparable with CHABADA. The main strength of CHABADA over other approaches is that it processes very lightweight information -- app description and API usage to flag an app as malware or not. However, if the description of the app changes and/or unavailable, the information that CHABADA has for that app is no longer valid. 

For example, CHABADA used `\textit{London Restaurants \& Pubs~+}' 
app as an example app to illustrate the false advertising and other questionable behavior unfortunately common to many published mobile apps. They identified that this Android app is possibly a malware and certainly has some unexpected behaviors. Finally, they placed it into the `TRAVEL\_AND\_LOCAL' category with a false malicious flag.  
Currently, this app is unavailable in the Google Play Store. It means that the description and screenshots for this app are also unavailable. However, we are interested in reclassifying \textit{'London Restaurants \& Pubs +'} in the current environment for extended decision making research purposes, e.g., Why might the assumption about malicious behavior become false? What are its most common characteristic features that match other malicious apps in the category? What are patterns in its data or API usages? Why was it categorized as 'TRAVEL\_AND\_LOCAL' rather than a SHOPPING (restaurant) or CASUAL (food)? 
However, meta-data and description based classification approaches are inferior for such extended decision making, an intrinsic limitation of the techniques~\cite{pandey2019perception}. 

It is also highly probable that the descriptions for an app published in 2014 are unavailable or have significantly changed by 2021. Also, app descriptions suffer all the usual natural language issues, including ambiguity, incompleteness and incorrectness, and malicious disguise of true app behaviors as they are mostly authored by humans. A possible solution to these problems could be to utilize only the data that exists inside the app for classification. There are a couple of potential arguments against this claim e.g., app code and GUI's also evolve over time, but as long as we have the app (apk file for Android app), it will remain valid and is reproducible.
\clearpage
\begin{table}[h!]
\centering
\caption{Processed data and REACT similarly score for "\emph{London Restaurants Bars \& Pubs +}" app}
\label{tab1}
\begin{tabular}{|>{\RaggedRight}p{0.07\textwidth} |>{\RaggedRight} p{0.38\textwidth} | } \hline

\textbf{Method names} & bulk compass rout \\\hline

\textbf{GUI text} & Null\\\hline

\textbf{XML data} & accur afghan african albanian alert american apart argentinian armenian asian australian austrian barbecu beach belgian bike bistro brazilian breakfast bridg british build burmes calcul cambodian canadian caribbean casino center chilean chines cinema club coffe colombian comput convent cream creperi croatian cuban czech dessert dinner distanc durat dutch ecuadorian educ egyptian elev english enhanc ethiopian european failur fast food fountain french garden german greek health histor hous hungarian imperi indian indonesian interact intern irish isra italian jamaican japanes keep korean landmark layer lebanes librari lodg london malaysian mediterranean metric mexican moment mongolian monument moroccan museum navig near neighborhood nepali order organ pakistani park pasta perform persian philippin pizza poi polish portugues proxim recommend religi restaurantsamppub romanian room rout russian satellit scottish seafood site spanish sport star streetview sushi swedish swiss taiwanes thai theater theme these tibetan tour tourist traffic transport tunisian turkish unit vegetarian vietnames walk waterfal\\\hline

\textbf{Highest similarity} &     Topic: T12 (91.274\% out of 100\%) 
                      \newline Corresponding category: Restaurant \& Non-Malware \\ \hline
\textbf{2$^{nd}$ Most similarity} &      Topic: T3
                        (6.725\% out of 100\%)
                 \newline   Corresponding category: Travel \& Non-Malware \\ \hline
\textbf{3$^{rd}$ Most similarity} & Topic: T32
                        (1.379\% out of 100\%)
                 \newline   Corresponding category: Entertainment \& Non-Malware \\ \hline

\end{tabular}
\end{table}

To address these issues we tried a novel \textbf{R}everse \textbf{E}ngineering-based \textbf{A}pproach to \textbf{C}lassify mobile app based on \textbf{T}he data that exists inside the app, called \textbf{REACT}.  Initially, REACT collects an apps' apk file from the existing app repositories and extracts all method names used in the raw code, XML data values, and GUI strings. These data are then processed to prepare a clean dataset. The clean dataset is the final input that is used in topic modeling for final classification decision. Following example clarifies how  REACT will remain valid and is reproducible in absence of app description.

We found the apk of this `\textit{London Restaurants \& Pubs~+}' app in AndroZoo server~\cite{AndroZoo}. 
Then, REACT extracted the data as mentioned above and processed it. The raw data extracted for this app is added in Appendix~\ref{Appedix-A-Chabada}. The processed data and REACT's most frequent similarly classified scores with at least 1\% contribution for this app is shown in Table~\ref{tab1}. From Table~\ref{tab1}, we found that the extracted and processed data for this app mostly contribute to Topic No. T12, which our REACT tool classifies as a Restaurant app and not malware. In an initial look to the process data, its looks more like a list of languages what is expected to be found in the XML for multi-language support app. However, the extracted words, `dinner, coffee and vegetarian' make a combined contribution to the country names and reveal it is a restaurant apps. 
In other words, meta-data based classification approaches\footnote{Throughout the paper, we consider the existing works \cite{Chabada}, \cite{AppClassfyMetaData} and \cite{AppClassfyMineText} as meta-data based approaches.} are not applicable to `\textit{London Restaurants \& Pubs~+}' in the current environment, whereas REACT is. 

However, our overall evaluation results show some limitations in our REACT approach and its current implementation compared to other existing approaches. To this end, this paper discusses the root cause of these failures, the key lessons learned, and some enhancement ideas for REACT. The work makes the following key contributions:

\begin{itemize}
\item We identified the point of failure of the existing app description and meta data-based classification schemes. We then designed a new approach (REACT) to try and overcome these limitations. REACT extracts and processes GUI strings, XML data values and method names from raw app code and categorise apps based on the similarity of extracted features using topic modelling results. 

\item To evaluate REACT we constructed a clean testing dataset for 19,766 Android apps that were previously used in the CHABADA work. We addressed two limitations with the original dataset. 
We also made our REACT dataset and tools publicly available in~\cite{Tools-Dataset-ICPC-2021-GITHub} that are reusable. 

\item While our final results show REACT is not as promising as we hoped for similar feature based app grouping, we learned several lessons that are discussed for the research community. 
   
\item We discuss how our reconstructed testing datasets and tools can be enhanced in the future to identify potentially malicious behavior of apps as well as to serve other app classification purposes.
\end{itemize}

The rest of the paper is organized as follows: Section~\ref{Sec2_CHABADA} presents the detailed theory, design and working procedure of the proposed REACT method. Then, Section~\ref{Sec3_Chabada} presents the research questions and corresponding evaluation results. The threats to validity, the key lessons we learned from developing REACT and the future works are discussed in Section~\ref{Sec4_CHABADA}.  Finally, brief related works and a set of conclusions are presented in Sections~\ref{Sec5_CHABADA} and \ref{Sec6_CHABADA}, respectively. 
 
\section{Our REACT Approach}
\label{Sec2_CHABADA}
\begin{figure} [h]
\includegraphics[width=.495\textwidth]{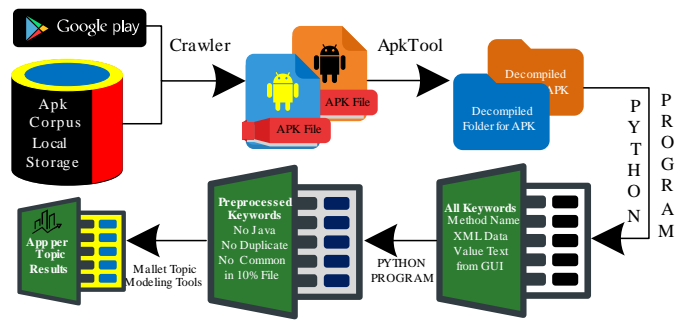}
  \caption{REACT Workflow}
  \label{Fig1}
\end{figure}

Key components of our proposed REACT approach are shown in Figure~\ref{Fig1}. The working procedure of the REACT can be divided into the following four stages: (i)~Android app collection and decompilation, (ii)~Feature extraction and cleaning, (iii)~Data preprocessing (iv)~Topic Modeling. 

\subsection{App Collection and Decompilation}
\label{App Collection and Decompilation}
We collected \emph{packageName} (\emph{appID}) and \emph{versionCode} for CHABADA apps\footnote{In this paper, we named all the Android apps that have been examined in CHABADA approach~\cite{Chabada} as CHABADA apps.} to search corresponding apk files in existing cloud repositories.  We tried to find the apk files with exact \emph{appID} and \emph{versionCode} first, then \emph{appID} and nearest \emph{versionCode}, and finally only \emph{appID} without any \emph{versionCode}. We  found 12,677 apps that exactly matches the \emph{versionCode} of CHABADA and 7,089 apps with a different \emph{versionCode}. However, we could not find apk files for 6,527 apps in existing cloud repositories (Play Store, AndroZoo, and third party cloud stores). The Crawler, python matching tool and dataset used for this process is available in~\cite{Tools-Dataset-ICPC-2021-GITHub}, where \texttt{Collected Apks Step 1 Output.xlsx} file summarized the results. Then we decompiled each apk file using the Apktool~\cite{apktool} because it can decode resources to nearly original form. Decompilation turns an apk file into a directory that contains all app codes and metadata. We stored decompiled app data for further processing. We are not sharing apk files or decompiled app directories as these uses several hundred GB of disk space. 

\IncMargin{.0in}
\begin{algorithm} [h!]
\SetKwData{Left}{left}\SetKwData{This}{this}\SetKwData{Up}{up}
\SetKwFunction{Union}{Union}\SetKwFunction{FindCompress}{FindCompress}
\SetKwInOut{Input}{Input}\SetKwInOut{Output}{Output}
\Input{A Decompiled apk directory}
\Output{Method named from all smali files}
\BlankLine
\Begin{
\For{j $\leftarrow$ all \textbf{*.smali} Files  \textbf{in}
     appDirectory}{
     f = open the file$_j$ and resolve encoding\\
    \For{k \textbf{in} f}{
        \If{match regular expression ``(.*)\.method(.*)" \textbf{in} k}
            {\textbf{return} characters after opening parenthesis \textbf{until} closing parenthesis $\rightarrow$ method\_name}
    } 
}
\For{l $\leftarrow$ allWord \textbf{in} method\_name}
    {separate the \emph{words} based on \textbf{isUpper}  function and \textbf{return} \emph{words}
    }
\textbf{remove} {all java keyword and single length characters sperate by space(s)} \textbf{from} \emph{words}\\
\textbf{return} \emph{words} \textbf{as} \emph{final keywords for method\_name}\\
}
\caption{REACT algorithm for method named extraction}
\end{algorithm}
\DecMargin{.0in}

\subsection{Feature extraction and cleaning}
In the second processing step, REACT parses three types of information from decompiled app directories: (i)~the entire directory of smali files to extract all method names, (ii)~string.xml file to extract all XML data values, (iii)~every image files from entire decompiled app directory to extract the GUI label text used in the app. The idea is to use these sources to identify distinct and similar representing phrases (features) of an app, and a group of apps at large.

While parsing smali files, our supposition is that developers generally use method names with component text related to its functionalities. For example, suppose a developer uses a method named `\texttt{getBirthData}'. In that case, we naturally expect that this method would collect the date of birth from the user, a database, or is elsewhere stored or calculated. Along similar lines, we expect developers to use words and phrases in the raw-code, data values or in GUI strings of an app that appropriately represent the key features/sub-features of apps. For example, an app asking the user for a birth date will likely have a GUI data selector with label `\texttt{birthDate:}', and/or an XML meta-data resource label e.g. `\texttt{birthDateValue}'. Hence, we expect that the extracted XML data values and GUI text labels will increase our set of distinct keywords indicative of key app features. 
Algorithms 1 and 2 show our data extraction process for method names, and XML data and GUI text values, respectively.

In Android smali files, method names start with a `\texttt{.method}' and the actual method name is contained inside parentheses. We match this through a regular expression `\texttt{(.*) ˙ method(.*)}' for all the smali files available in the app directory, shown in lines 4 and 5 of Algorithm 1. We then remove following set of Android keywords as they rarely contribute to app classification: \\

\noindent\fbox{%
    \parbox{.4725\textwidth}{%
        {\$, \_, -, $<$clinit$>$, $<$init$>$, , abstract, assert, boolean, break, bridge, byte, case, catch, char, class, const, constructor, continue, create, declared, default, do, double, else, enum, execute, extends, false, final, finally, float, for, get, goto, has, if, implements, import, instanceof, int, interface, iterator, long, native, new, next, null, on, package, private, protected, public, return, run, set, short, static, super, switch, synchronized, synthetic, this, throw, throws, to, transient, true, try, value, void, volatile, while, All, Button, Click, Down, Drawable, Drop, From, Icon, Item, Layout, Menu, Next, String, Title, To, Value, View}. 
    }%
}
\newline

Then, we separate the extracted phrases to word as per CamelCase convention. REACT inherently handle the snake\_case coding convention as it removes underscore before the phrase-to-word extraction step. A similar regular expression is used to extract XML data values (shown in lines 5 of Algorithm 2).   We also parse the entire app directory to extract image data and check which files can be opened using the Python `\texttt{Image.open()}' function. We then extract text from the images and remove all single length characters, shown in lines 7-9 of Algorithm 2. The extracted data are stored in \texttt{Dataset\_1.xlsx} excel file that has 5 columns and 19,767 rows, where each row contains the data for an app. The cells of each row contains app's SHA256, package name, extracted method name, XML data value, GUI text data (in that order).

\IncMargin{.0in}
\begin{algorithm} [t]
\SetKwData{Left}{left}\SetKwData{This}{this}\SetKwData{Up}{up}
\SetKwFunction{Union}{Union}\SetKwFunction{FindCompress}{FindCompress}
\SetKwInOut{Input}{Input}\SetKwInOut{Output}{Output}
\Input{A Decompiled apk directory}
\Output{XML data value and texts for image files}
\BlankLine
\Begin{
\For{j $\leftarrow$ all \textbf{string.xml} Files  \textbf{in}
     appDirectory}{
     f = open the file$_j$ and resolve encoding\\
    \For{k \textbf{in} f}{
        \If{match regular expression ``$>$([\char `\^ $>$]*)$<$\/" \textbf{in} k}
            {\textbf{remove} all single character from $k$\\
            }
    } 
}
\For{l $\leftarrow$ all \textbf{image} Files  \textbf{in}
     appDirectory}{
     f = open the file$_l$, extract text and resolve encoding\\
     \textbf{remove} all single character from $l$\\ }
\textbf{return} \emph{k} \textbf{as} \emph{final keyword  for XML\_Data\_Value} and  \emph{l} \textbf{as} \emph{final keyword for GUI\_text }
}
\caption{REACT algorithm for XML data value and GUI text extractions}
\end{algorithm}
\DecMargin{.0in}

\subsection{Data Preprocessing and Dataset Preparation}
\label{Data Preprocessing and Dataset Preparation}
Thirdly, REACT sequentially clean and process the data to remove uninteresting information. For example, initially REACT sort the data and remove duplicate keywords because frequency of word add no useful information to REACT classification in later stage for the similar reason explained in~\cite{freq-no-useful}. Then, all keywords that have less than four characters are removed because these single, two or three letters keywords are mostly truncated variable names and XML tags that we found are not assisting in app classification. 

REACT counts the frequency of each individual keyword and removes all keywords that have more than 10\% support (the occurrence frequency in percentage). This is because these keywords will not help us to uniquely classify an app or app group at large as mentioned early. Then, REACT applies Stemming and Lemmatization to all the remaining keywords based on the Python library implementation~\cite{stem2}. The reasons to do the Lemmatization and apply Stemming in REACT Data are: (i)~they help to achieve the root forms of the extracted keywords, (ii)~they can  improve the precision by estimating word similarity during the final outcome, and (iii)~sorting of keywords and removing duplicates from the list makes our dataset more compact and hence boosts  computational performance. We initially tried both Porter and Lancaster stemming methods. We decided to use the Porter stemming algorithm since the Lancaster algorithm sometimes changes the meaning of our actual keywords. More detailed discussion on Lemmatization and Stemming is beyond the scope of the paper, and the interested reader can see~\cite{stem3}. Finally, REACT removed any duplicate keywords that arise due to stemming. Algorithm 3 shows these sequential preprocessing steps. 

The preprocessing greatly reduces the size of the dataset. For example, before the preprocessing the average number of words for method names is 1,915, XML data value is 421 and words extracted from image files is 85. After preprocessing these numbers become 61, 70, and 26, respectively. We also deleted 345 rows from previous dataset (corresponds to 345 apps) because they have (i) less than 10 keywords, (ii) at least 10\% non-English words, or (iii) $\geq$51\% encrypted data. All these together, helps us to prepare an appropriate clean dataset (\texttt{Dataset\_2.xlsx} in our shared replication package~\cite{Tools-Dataset-ICPC-2021-GITHub}).

\IncMargin{.0in}
\begin{algorithm} [t]
\SetKwData{Left}{left}\SetKwData{This}{this}\SetKwData{Up}{up}
\SetKwFunction{Union}{Union}\SetKwFunction{FindCompress}{FindCompress}
\SetKwInOut{Input}{Input}\SetKwInOut{Output}{Output}
\Input{Dataset that contains method name, XML data value and GUI text}
\Output{Preprocessed and compact dataset}
\BlankLine
\Begin{
\For{j $\leftarrow$ all data \textbf{in} dataset}{
        \If{j \textbf{contains} upper case character}
           {\textbf{convert} \emph{data}$_j$ \textbf{into} lower case}
         \textbf{sort} \emph{j} and \textbf{remove} duplicate \textbf{from} \emph{j}\\
         }

\For{j $\leftarrow$ words \textbf{in} dataset}{
        \If{word\_length$_j$ $\leq$ 3}
           {\textbf{remove} word$_j$}
        }
\For{j $\leftarrow$ words \textbf{in} dataset}{
         \textbf{lemmatize} word$_j$
         }

\For{k $\leftarrow$ words \textbf{in} dataset}{
          \textbf{apply porter stemming} word$_k$\\}

          \textbf{sort} and \textbf{remove} duplicate words \textbf{from} \emph{dataset}\\

\For{j $\leftarrow$ words \textbf{in} dataset}{
         \If{frequency$_j$ $\geq$ 10\%}{
            \textbf{remove} \emph{j}
        }
}

\textbf{return} \emph{data} \textbf{as} \emph{preprocessed keywords}\\
}
\caption{REACT algorithm for data preprocessing}
\end{algorithm}
\DecMargin{.0in}

\subsection{Topic Modelling}
\label{MALLET-Class}
REACT uses topic modeling algorithms to discover the abstract `topics' and identify the hidden semantic structures. It uses the clean and preprocessed data prepared in the previous step to form the `topics' from similar keyword clusters. In REACT, we tried two most popular topic modelling techniques (i) \textbf{L}atent \textbf{D}irichlet \textbf{A}llocation (LDA) and (ii)  \textbf{MA}chine \textbf{L}earning for \textbf{L}anguag\textbf{E} \textbf{T}oolkit (MALLET). LDA is a generative statistical model. It allows sets of observations to be explained by unobserved groups and explains why some data parts are similar. MALLET implements LDA but uses the Gibbs sampling methods. MALLET is a Java-based package put out by UMASS Amherst for statistical natural language processing, classification, clustering, and information extraction. LDA uses a Variational Bayes sampling method, whereas MALLET implements LDA but uses the Gibbs sampling methods. Thus, LDA is faster but less precise than MALLET’s sampling~\cite{TM2}.

\begin{figure*}[ht]
  \centering
  \includegraphics[width=.975\textwidth]{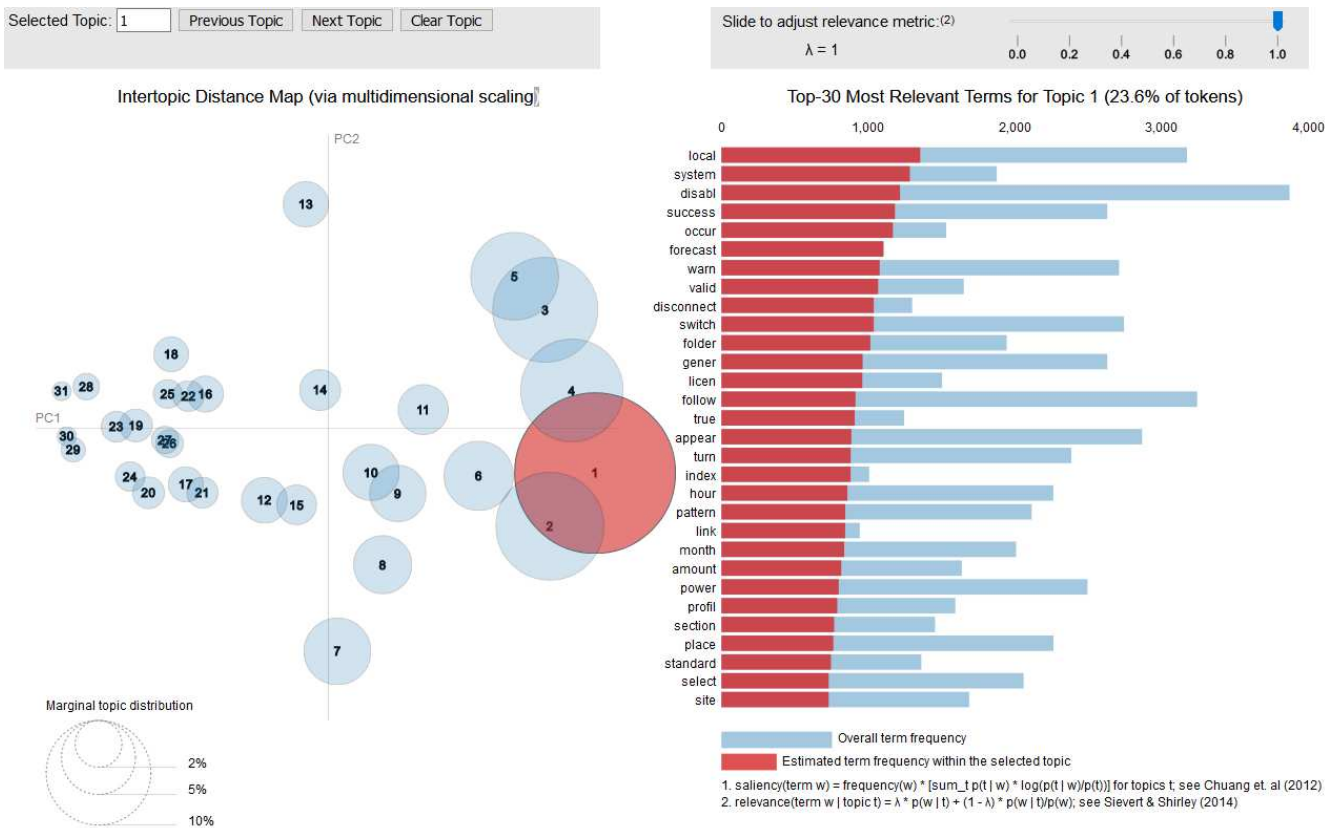}
  \caption[]{LDA Outputs for all Topics in REACT}
  \label{FigLDA1}
\end{figure*}

The output for REACT LDA model with 31 topic based on pyLDAvis package’s chart~\cite{LDA_output} is shown in Figure~\ref{FigLDA1}, where each bubble on the left-hand side plot represents a topic. The larger the bubble is, the more prevalent is the topic. In this analysis, we aimed to identify distinct topic from  non-overlapping bubbles scattered throughout the chart and then evaluate its functionalities. However, as shown in the Figure~\ref{FigLDA1}, we failed to do so appropriately. Therefore, we tried MALLET that to identify more concrete outcome using our clean dataset following the Shawn Graham et al. method~\cite{MALLET}.  This scheme requires a separate text file for each app that contains all its extracted text keywords, which done through a VB script (\texttt{rowAsTxtFile}). Then, MALLET is applied and it produces two output files \texttt{tutorialKeys.txt} and \texttt{tutorialComposition.txt}. We stored these results in \texttt{MALLET Result.xlsx} dataset that have two sub-sheets. The first sheet contains the Topic No, Dirichlet parameter for the topic, and the most representative keyword for that topic. In sheet 2, there are 32 columns and 19,442 rows. The first column represents the app row from the dataset, and the next 31 columns represent contributions to each topic from this particular app in percentages.
We then identified four most contributory topics for each row (correspond to each app). Here, we consider the highest contributory topic as the `primary topic' for an app. If a topic contribution is less than 1\%, we discard it from the contributory list. Topic  No 1 to 31 are named as shown in the following box and in that order: 
\\

\noindent\fbox{%
    \parbox{.4725\textwidth}
        {{Personalization, Cards, Education, Libraries\_and\_Demo, Lifestyle, Tools, Medical, Music\_and\_Audio, Sports, Arcade, Transportation, Business, Non-Match (392 Apps), Communication, Casual, Brain, Health\_and\_Fitness, Finance, Productivity, Books\_and\_Reference, Racing, Photography, Entertainment, Media\_and\_Video, Comics, Weather, Travel\_and\_Local, Sports\_Games, News\_and\_Magazines, Shopping, Social}
        }
    } 
\\
\\
Our evaluation also found that more than 67\% apps belong to at most two topics with $\geq$39\% contribution. We compare the topic similarity based on the most frequent match in the final step. A more detailed discussion about our findings and results are presented in the following sections. 
\section{Comparative Analysis}
\label{Sec3_Chabada}

This section presents our evaluation results of REACT. We wanted to answer the following three key \textbf{R}esearch \textbf{Q}uestions (RQ) in these evaluations:

\begin{description}
\item [RQ1] \textbf{How does REACT classify mobile apps without app descriptions?} To evaluate this question, we present detailed comparative results and analysis in Sec.~\ref{REACTRQ1}. We also discuss with examples why some of our results were not as promising as we hoped. Some future research ideas are also introduced here.
\\
\item [RQ2] \textbf{How can REACT datasets and tools be enhanced to detect anomalies?}  
In other words, does combining the keywords extracted from XML data value, text from the image files and method names from smali files of Android apps lead to the construction of a dataset that can be enhanced in the future for malware and anomaly detection similar to the existing meta-data based approaches? This question is evaluated in Section~\ref{REACTRQ2}.\clearpage

\item [RQ3] \textbf{How reliable and scalable is our proposed REACT approach?} 
i.e., Do the REACT algorithms and tools perform equivalently for a new input set? We answer this research question in  Section~~\ref{REACTRQ3} considering a new set (not examined previously) of 5,000 Android apps.
\end{description}

\subsection{RQ1 - App Classification in the Absence of Descriptions}
\label{REACTRQ1}
REACT classifies Android apps without using any app description or screenshot, unlike existing meta-data based approaches. We  discussed in Sec.~\ref{Sec1} that the CHABADA approach has several advantages over other existing techniques. We thus wanted to evaluate the REACT tool-set on the same set of CHABADA apps. We were successful in collecting 75\% CHABADA apps (19,766 Android apps out of 26332 apps). It is acceptable in the sense that eight years have passed since CHABADA apps were examined. We also discussed other possible reasons for the inaccessibility of the remaining 25\% apps at the very beginning of the paper.

\subsubsection{Classification Result Analysis in App Categorization}
We tried to identify distinct topics in REACT using LDA topic modeling technique. However, overlapping results cause difficulties to do this appropriately, and hence we tried MALLET. In Section~\ref{MALLET-Class}, we present the topic name and corresponding categories for MALLET based classification for each topic. The similarity result between REACT and the CHABADA methods are shown in Table~\ref{tab5} in terms of most frequent matches. From this table, we find that the Topic~26 has the highest similarity 78.35\% in both prototypes that REACT categorized as `\emph{WEATHER}'. The 78.35\% similarity means that the REACT assigns 78.35\% of the apps classified in `\emph{WEATHER}' category that were also categorized as `\emph{WEATHER}' in CHABADA. However, the average classification similarity between REACT and CHABADA is not much high, some possible reasons for this are: (i)~CHABADA author gave custom names to the categories, there is no guarantee that the topics we deal with are the same. (ii)~Overlapping inputs for different app types that go to topic modeling cause two completely different apps classified in the same category, (iii) We only consider the most frequent match during classification, (iv) The number of topic might be incorrect in both REACT and CHABADA approaches.

\begin{table}[h!]
\footnotesize
\caption{Similarity results between REACT and CHABADA approaches for app categorization in terms of the most frequent match}
\label{tab5}
\begin{tabular}{||c|c||c|c||c|c||}
\hline
\textbf{Topic} & \textbf{Similarity} & \textbf{Topic} & \textbf{Similarity} & \textbf{Topic}                 & \textbf{Similarity }              \\ \hline
T$_1$         & 8.28\%     & T$_{12}$      & 31.19\%    & T$_{23}$                      & 2.41\%                   \\ \hline
T$_2$         & 12.19\%    & T$_{13}$      & NA         & T$_{24}$                      & 65.12\%                  \\ \hline
T$_3$         & 10.43\%    & T$_{14}$      & 8.80\%     & T$_{25}$                      & 5.23\%                   \\ \hline
T$_4$         & 5.76\%     & T$_{15}$      & 19.42\%    & T$_{26}$                      & 78.35\%                  \\ \hline
T$_5$         & 5.99\%     & T$_{16}$      & 19.46\%    & T$_{27}$                      & 7.75\%                   \\ \hline
T$_6$         & 6.68\%     & T$_{17}$      & 10.41\%    & T$_{28}$                      & 28.32\%                  \\ \hline
T$_7$         & 48.70\%    & T$_{18}$      & 2.53\%     & T$_{29}$                      & 13.82\%                  \\ \hline
T$_8$         & 12.43\%    & T$_{19}$      & 9.55\%     & T$_{30}$                      & 12.80\%                  \\ \hline
T$_9$         & 3.00\%     & T$_{20}$      & 14.10\%    & T$_{31}$                      & 62.85\%                  \\ \hline
T$_{10}$        & 35.98\%    & T$_{21}$      & 7.18\%     & \multirow{2}{*}{Average} & \multirow{2}{*}{19.48\%} \\ \cline{1-4}
T$_{11}$        & 24.93\%    & T$_{22}$      & 10.78\%    &                          &                          \\ \hline
\end{tabular}
\end{table}
\begin{table}[t]
\centering
\footnotesize
\caption{Processed data, similarly scores and corresponding categories for four Android apps}
\label{tab22}
\begin{tabular}{ |>{\RaggedRight}p{.475\textwidth}|} \hline

\textbf{App package ID:} com.dipandro.nqlv1 \newline
\textbf{Extracted keywords (Method name; XML data; GUI data):} edstart facebook like question quiz;	anatomi basal brain cerebellum cerebr complimentari cortex disord excerpt ganglia ganglion lite neurolog neuroquizact okay physiolog question quiz stem volum; null	
\\
\textbf{Top similarity match in REACT:} T7	52.23\%,	T28	28.42\%,	T31	12.39\% \newline

\textbf{REACT Category:} Medical  \newline
\textbf{CHABADA category:}  Medical \\
\textbf{Human-judged category:} Medical\\ 
\hline

\textbf{App package ID:} com.hzwp.mtFlowers \newline
\textbf{Extracted keywords (Method name; XML data; GUI data):} angl brows coverflow dcard fling fold folder judg messg quit saveto threadsleep thumb unzip wallpap xmlpar xmlparser;	choic collect effect excel flower follow jjwallpap larg moment nice notic origin path prev rateapp reach resourc rotat save score softwar veri warn zoom; 	cocotocti reaossssssss rrroe
 \\
\textbf{Top similarity match in REACT:} T24	50.27\%	T28	43.72\%	 \newline
\textbf{REACT category:} Media\_and\_Video  \newline
\textbf{CHABADA category:}  Media\_and\_Video\\
\textbf{Human-judged category:}  Media\_and\_Video \\ 
\hline

\textbf{App package ID:} com.gregorbrett.sapsearcher 	 \newline
\textbf{Extracted keywords (Method name; XML data; GUI data):} desc exampl favorit modul tcode; favorit saptcodesact searcher; brows descript favorit modul name sapsearch search transact\\

\textbf{Top similarity match in REACT:} T31	71.31\%	T9	11.86\%	T7	6.11\%	T11	5.97\%
  \newline
\textbf{REACT category:} Social \newline
\textbf{CHABADA category:} Business  \\
\textbf{Human-judged category:} Tools   \\ 
\hline

\textbf{App package ID:}com.holfeld.germanwordsfree  	 \newline
\textbf{Extracted keywords (Method name; XML data; GUI data):} actual assign declar exclud exclus execut expos owner pring serial; aecddaf german germanwordsfre marketdetailsidcomholfeldgermanword word; bcplyst calendar german lama parri shee speak word wwwholfeldappscom \\

\textbf{Top similarity match in REACT:} T6	42.53\%	T14	30.88\%	T20	15.57\%	T3	7.77\%
 \newline
\textbf{REACT category:} Tools \newline
\textbf{CHABADA category:} Business   \\
\textbf{Human-judged category:} Education   \\ 
\hline

\textbf{App package ID:} com.brainvision.foodchecklist \newline
\textbf{Extracted keywords (Method name; XML data; GUI data):} blacklist food whitelist; carboh checklist food occur protein restart; null \\
\textbf{Top similarity match in REACT:} T7	45.96\%,	T24	37.44\%, 	T28	9.98\% \newline
\textbf{REACT category:} Medical\newline
\textbf{CHABADA category:}  Health\_and\_Fitness  \\
\textbf{Human-judged category:} Health\_and\_Fitness \\ 
\hline
 
\textbf{App package ID:} com.CPGSGoalieStats	 \newline
\textbf{Extracted keywords (Method name; XML data; GUI data):} ppon pass season shot side singleton spinner team; addit comment contain exampleemailcom goaliestat goaliestatsact letter; heme keema undo\\

\textbf{Top similarity match in REACT:} T31	63.79\%	T28	32.14\% \newline
\textbf{REACT category:} Social \newline
\textbf{CHABADA category:}  Sports\\
\textbf{Human-judged category:}  Sports \\ 
\hline

\end{tabular}
\end{table}

To further evaluate the classification results, we compare REACT similarity unitedly with human-judge categories and CHABADA prototype for a random set of 250 apps. An example comparison for six apps is presented in Table~\ref{tab22}. The first two rows of Table~\ref{tab22} show that the classification results are identical in REACT, CHABADA, and Human-judged approaches. However, the third and fourth rows show that the `\textit{com.gregorbrett.sapsearcher}' and `\textit{com.holfeld.germanwordsfree}' apps are dissimilar in all three schemes. This result implies that we might need more data from the app itself for proper classification.
We also find that the identical results for the former case were possible because (i) extracted and processed features are distinct from other app groups by a good margin, (ii) very few overlapping keywords are extracted, (iii) final input that goes to topic modeling does not include any encrypted or obfuscated data.  Moreover, the last two rows of Table~\ref{tab22} shows that the similarity results for the  \textit{com.brainvision.foodchecklist}\, and\, \textit{com.CPGSGoalieStats} \, apps are different in REACT compared to CHABADA but are identical to human judged categories. This result justifies our claims about custom names of categories for CHABADA apps discussed at the beginning of this section.

In this manual evaluation, we also found (i)~37.2\% app is in the same category in all three approaches, (ii) additional 26.4\% are same in REACT and Human-judge category but different in CHABADA, (iii) remaining apps are different in all three evaluation. In the second result, 40.65\% apps (37 out of 91) are of a different version than what was evaluate in CHABADA. Hence, we believe the issue of `how does app functionality change with version change?' could be an interesting future research question to investigate.

\subsubsection{Classification Result Analysis for Historical Apps}
We define ``historical app" as a mobile app whose description is either unavailable or changes substantially over time. Since meta-data based approaches begin classification from modelling the topic or training the data from app descriptions, these are either completely inapplicable or will produce incorrect results for historical apps. Consider a historical app that was previously classified as malware or assigned to a specific topic by a meta-data based approach. Now, we are interested to evaluate its current state. The meta-data based prototypes rarely help in this case, because the app description might be unavailable or substantially changed. However, REACT is reliable and scalable in such scenarios as it analyzes the extracted data (static) that is available inside the app, i.e., as long as we have an apk file of an app, REACT can extract data and process it appropriately.

\subsection{RQ2 - Extendability and Reproducibility}
\label{REACTRQ2}

\subsubsection{Tool Reuse}
The python, R and bash scripts based CHABADA tools are currently inaccessible. However, re-implemented CHABADA techniques have been open-sourced since 2016 but not adequately publicized\footnote{Available in Git-based source code repository (Bitbucket) as Untitled project in \url{https://bitbucket.org/gorla/smapper-imdea-anomalydetection}}, especially not under the name of CHABADA. Even then, it requires considerable effort and time to reproduce this work, which is also true for other meta-data based approaches. To counter such argument in REACT, we make REACT tool-sets publicly available for reuse, reproduction, comparison, and extension in~\cite{Tools-Dataset-ICPC-2021-GITHub}. 

\subsubsection{Data-set Reuse}
The authors of CHABADA  mentioned that they prepared a dataset `with the exact data used in CHABADA, including app names, descriptions, permissions, API usage and other metadata'. We have found that these data are mostly represented as flagged, i.e., true(1) or false(0). Similarly, the publicly available dataset from other meta-data based prototypes also contains processed data and final results. Hence, it is tough to reuse these datasets to enhance the works themselves, e.g., modifying the analysis approaches for other decision-making research except what was examined. 

To address the reuse issues REACT, we make all datasets publicly available in~\cite{Tools-Dataset-ICPC-2021-GITHub}. For example, `\texttt{Dataset\_1.xlsx}' file contains the initial extracted data, i.e., method names, XML data value, and text data pulled from image files. This file also contains the app's package name and SHA256 at the beginning of each row to identify an app uniquely. The `\texttt{Dataset\_2.xlsx}' file contains preprocessed and clean data, whereas `\texttt{Dataset\_3.xlsx}' contains a summary of topic modeling results and final data that goes to the processing. The detail of topic modeling results is accessible from `\texttt{Dataset\_4.xlsx}' file. Researchers, developers, and app market analysts can use these resources for possible research enhancement, replication, and reproduction. 

\subsubsection{Enhancement to Detect Anomalies}
The REACT datasets can be used to detect anomalies or malicious app behavior. Overall, if the extracted keywords contain obfuscated data, encrypted data and contribute to multiple topics by $\geq8\%$ in each, REACT considers this app as a high potential candidate for being malicious. One could argue that, obfuscation is nowadays very widespread and developers are encourages to obfuscate app code by simply checking a flag in the build process. We counter this as `Our claim for anomaly detection will be true only when obfuscation is noticed along with encrypted data and it contribute to multiple topics (more than 8\% in each topic in the final modeling result)'. 

To further explain this, we manually checked the first 25\% apps that were marked as malicious in the CHABADA. We only found apk files for six apps. The corresponding analysis results are shown in Table~\ref{tab44}. From this table, we find that four apps (row 1, 2, 3, and 6) can also be marked as malicious as per REACT analysis. For example, the app `\emph{net.bible.android.activity}' was marked as malicious by the CHABADA prototype. In REACT, it contributes to multiple topics where the four most frequent contributory topics are T12(30.95\%), T24(23.15\%), T22(16.77\%) and T4(10.99\%). It also contains obfuscated data in its method names and XML data values. Some encrypted data was also found in its raw data. Hence, we indicate that it is highly possible that it functions maliciously. The app shown in rows 4 and 5 of Table~\ref{tab44} cannot be identified as malicious thorough REACT, because there is no encryption or obfuscation. However, the apk files of these two apps that are evaluated in REACT are different versions than what was evaluated in the CHABADA prototype. Therefore, its malicious behavior may have been changed (removed) in the new versions. Moreover, it was not our primary aim to detect the anomalies or malicious apps in REACT, rather, we argue that the proposed REACT dataset can also be used (enhanced) to serve a similar purpose to the previous meta-data based approaches.

\begin{table}[h!]
\centering
\caption{REACT results for first 25\% app that were marked as malicious in CHABADA}
\label{tab44}
\begin{tabular}{|>{\RaggedRight} p{0.47\textwidth} | } \hline
\textbf{App package name:} net.bible.android.activity
\\
\textbf{App ID, Row no in CHABADA:} 38, 37
\\
\textbf{Results in REACT:} T12(30.95\%), T24(23.15\%), T22(16.77\%) and T4(10.99\%); Contains obfuscated data in method name and XML data value, encryption found in raw data. 
\\
\hline

\textbf{App package name:} com.nubee.coinpirates
\\
\textbf{App ID, Row no in CHABADA:} 48, 45
\\
\textbf{Results in REACT:} T9(55.84\%),	T31(20.44\%), T23(12.14\%), T28(10.73\%); Contains obfuscated data in XML data value, encryption found in raw data. 
\\
\hline

\textbf{App package name:} com.lonelycatgames.Xplore
\\
\textbf{App ID, Row no in CHABADA:} 62, 59
\\
\textbf{Results in REACT:} T24(70.26\%),	T27(11.85\%),	T8(8.99\%),	T28(5.35\%); Contains obfuscated data in XML data value, encryption found in raw data. 
\\
\hline

\textbf{App package name:} com.electricsheep.dj
\\
\textbf{App ID, Row no in CHABADA:} 66, 63
\\
\textbf{Results in REACT:} T15(71.08\%),	T28(9.44\%),	T24(6.64\%),	T10(6.59\%); Doesn't contain any obfuscated or encrypted data   
\\
\hline

\textbf{App package name:} com.reverie.game.toiletpaper
\\
\textbf{App ID, Row no in CHABADA:} 70, 66
\\
\textbf{Results in REACT:} T16(31.68\%),	T28(26.50\%),	T24(21.25\%),	T22(14.10\%); Doesn't contain any obfuscated or encrypted data   
\\
\hline

\textbf{App package name:} org.openintents.filemanager
\\
\textbf{App ID, Row no in CHABADA:} 71, 67
\\
\textbf{Results in REACT:} T24(87.80\%), T12(8.51\%); Contains obfuscated data in XML data value, encryption found in raw data. 
\\
\hline
\end{tabular}
\end{table}

\subsection{RQ3 - Scalability and Reliability}
\label{REACTRQ3}
RQ3 asks about our REACT tool performance for extracting, processing and outcome evaluation. To answer this question, we applied REACT to a new set of 5,000 Android apps that were not previously analyzed. We collected these apps from our previously collected corpus of 1 million Android apps. Due to the blind review process, server IP address and identification information is kept hidden. It is also impossible to share the apk files in our shared replication packages because they take over 200 GB storage. However, these do not hinder the understandably RQ3 analysis presented below.

The new 5000 apps were chosen as per the first alphabetic order of SHA256. Thus, app selection becomes random as per app ID (package name). We removed 130 poor quality apps from this list and then added 16 validated eHealth apps. The low-quality apps are so marked if their extracted data are thoroughly encrypted or not readable. The good eHealth apps are defined as per the MARS scale~\cite{MARS} with an average score over 3.0. The extracted data without duplicates for these 4,886 new apps are stored in `\texttt{Dataset\_5.xlsx}' file for further investigation. To assess the scalability of REACT, we tried to modify our python programs to extract the app name (title) rather than app package name from the apk file. Here, we found that we only need to modify the input file location and corresponding conditions, only four lines of python codes need changes. The final and processed data for these apps are stored in `\texttt{Dataset\_6.xslx}'. From this new dataset, REACT inherently identifies 2,128 apps as uninteresting apps (with obfuscated and encrypted data) that justify our claims on reliability. 

We ran this part of our experiments using the same REACT tool on a workstation with `Intel(R) Xeon(R) W-2175 CPU @ 2.50GHz, 32GiB System memory, and  Ubuntu SMP GNU/Linux kernel'. On average, each app extracted data from this new set requires 1,8884.50 bytes storage and 3.1863 seconds processing time compare to 4509.10 bytes and 5.8311 seconds in the previously examined dataset (19,766 apps dataset). Additional possible threats related to these analysis 
are discussed in the following section. 
\section{Discussion}
\label{Sec4_CHABADA}
In REACT, we reproduced an Android app classification scheme similar to existing meta-data based approaches, but without using human-authored app descriptions or dynamic data that changes over time and may be unavailable. Moreover, we prepared new clean datasets that are extendable to identify anomalies and malicious activity in mobile apps. We have made all of our datasets and tools publicly available in~\cite{Tools-Dataset-ICPC-2021-GITHub}. Detailed documentation for these datasets and tools are provided for future enhancement. 
In the following sections, we discuss several threats related to REACT evaluation results, some of which need further investigation. 

\subsection{Threats to Validity}
In REACT, we extract data from currently available CHABADA apps. We tried our best to collect the full set of apps but were only partially successful. We explained the reasons for accepting the partial collection at this stage in Sec.~\ref{REACTRQ1}. We also used a new set of 5000 apps to show that the computational resource requirement in REACT changes linearly, requires one-fourth of resources for one-fourth input size. In this regard, a potential future work could be evaluating REACT performance over larger app input size. 

Our proposed REACT approach is also subject to an additional construct and internal validity  threats. For example, we removed a set of ``uninteresting" apps (345) before the Stemming and Lemmatization steps. It helps us to prepare a clean dataset, which is explained in Sec.~\ref{Data Preprocessing and Dataset Preparation}. However, it is an intrinsic limitation of REACT, and it should perform equivalently even in the presence of uninteresting apps. Moreover, identifying patterns from extracted raw data (\texttt{Dataset\_1.xsls}), including the unrelated and uninteresting apps without app removal, is another interesting future research. We also present three research questions and explain our answers to these questions with examples. However, we were unable to avoid partial non-promising results for the first research question (classification). 

In addition to the above threats, there are two additional limitations in REACT. First, REACT has the same threats as existing static data analysis based decision making research tools i.e., ignoring run-time generated data. The dynamic analysis was not used in part so that replication can be carried out without concern about historical app behavior. Secondly, all REACT experiments, dataset preparation and result analysis were primarily based on examination by a single person, who is a co-author of this paper. Then, three renowned software engineering researchers cross-checked partial results. Hence, it still has a risk of producing bias results addressing the examiner and these expert needs only. To counter this issue, we made all extracted, preprocessed and final datasets publicly available along with our toolkit for others to examine, use and extend. 
\subsection{Key Lessons Learned} 
In REACT, we were only able to compare results currently available CHABADA apps. Apps no longer available in any public forum could not be compared with previous analysis approaches and tools. Similarly, apps which have evolved but the earlier versions are no longer available could only be analyzed using their later versions. This highlights that to do comparative app analysis with often highly evolving apps, careful curation of historical app data is needed. Below we summarized additional findings and lesson we learned. 

\textbf{\\Obfuscation:} In Android apps, data obfuscation and encryption is mostly done on method names and XML data values rather than GUI strings. For example, in our REACT dataset more than 95\% obfuscated and encrypted data lies in methods names and XML data values. Hence, appropriate data cleaning is required for any decision-making research that plans to use raw code and XML data.  
\\
\\
\textbf{Data Value Sharing:} Two quite different types of Android apps can share substantially similar set of XML data values. For example, two different types of apps `\textit{com.gaiaonline.mge}' (row 668 in REACT dataset) and `\textit{com.livewall.Galaxy}' (row 1070 in REACT dataset) share more than half of their extracted data. Even then REACT classifies the first one MUSIC\_AND\_AUDIO apps, whereas the second one as LIFESTYLE app. The different classification was possible due to the use of stemming and lemmatization, explained in Sec.~\ref{Data Preprocessing and Dataset Preparation}. The data similarity is also observed for other data type in REACT app sets. Therefore, we think that additional data type other these two will produce better classification results. In this regard, we collect all image files from the apps' decompiled apk and extracts all available strings (text) from these files.  However, it turns out that the data  pulled from image files is mostly uninteresting and rarely helps in decision making but requires more resources. 
\\
\\
\textbf{Combining App Source and Human Data:} Combining new app analysis data with manually extracted data could improve the classification of app groups. However, in such case the method would not be suitable for historical apps (defined earlier) as well as for more larger input set.  
\\
\\
\textbf{Analysis tasks:} REACT analysis largely depends on the outcomes from LDA and its implementation in MALLET, both of which are probabilistic. Therefore they returns different topics when performing repetitions on the same dataset i.e., stochastic in nature and is a limitation of REACT. However, REACT tools and datasets could be applied to other app  analysis tasks. An interesting future work is to identify outliers from each data type and then classify. In such case its performance can be evaluated over existing results. Another interesting work could be collecting a good number of authentic apps from Google play store and apply REACT tools to see how it performs.

\section{Related Work}
\label{Sec5_CHABADA}
Mobile app classification is a well-studied area in the literature~\cite{li2019rebooting, kong2018automated, li2017static, martin2016survey}. Some of the main reasons to automatically classify apps or to validate app classifications include: (i) Malicious activity detection or malware identification -- researchers compare an apps' expected behaviour to observed behaviour based on the app's classification \cite{Chabada,Ma_2015,FEIZOLLAH201522, yang2017characterizing, li2017understanding}.
(ii) Pattern recognition and to distinguish app relationships from a particular domain -- to help users for finding their favorite app based on retrieve pattern and compare similarities among different version of an app or repackaged apps to provide recommendation~\cite{LiSimilarities}.
(iii) Design and working procedure specification based on extracted features -- new features are recommended to help app developers discuss methodological approaches to app studies. These can be for different app settings and for drawing out common aspects, trends and directions to address open problems and challenges in app development~\cite{MArtinTSE}
(iv) More accurate classification to overcome the app store categorisation problems -- where problems with human-authored classifications are addressed by automatic scheme~\cite{TSE_miscatagorization,iTuneCategory},  
(v)~Automate app verification -- where classifications assist in the verification processes used~\cite{Chabada}~etc. 
A few of these existing approaches 
have challenges in terms of being reusable or reproducible, especially to deliberate app behaviors for a variety of classification purposes. The primary reason for this challenge is some of these data might be unavailable or changed, as we have already discussed in the previous sections. 
\section{Conclusion}
\label{Sec6_CHABADA}
In this paper, we instanced the point of failure for the existing app description and meta-data based mobile app classification methods. We then investigated a reverse engineering-based approach called REACT to avoid such failure and reproduce results using the data that exists inside the apps. To achieve these goals, we collected 19,766 Android apps that were examined in the well-known CHABADA prototype. We then reverse engineered these apps (apk files) to extract keywords from (i)~method names used in the raw app code, (ii)~XML data value, and (iii)~text from image files. The extracted data were then preprocessed, and a clean dataset was prepared. However, comparative analysis shows that the proposed scheme is not as accurate as we hoped, we discussed (i)~REACT's enhancement for mobile app anomaly detection analogous to the existing approaches, (ii)~REACT's scalability and reliability, evaluated through a new set (not examined previously) of 5,000 Android apps, (iii)~key lessons learned for future app analysis tasks, replicating prior work, and handling rapidly evolving app versions for analysis, and (iv) REACT's threats to validity with a set of recommendations. Moreover, we share the REACT tool-set and all datasets for further research purposes in~\cite{Tools-Dataset-ICPC-2021-GITHub}. 

\section*{Acknowledgements}
Shamsujjoha is supported by Monash International Tuition Scholarship, RTP Stipend, and CSIRO Data61 Top-up Scholarship for his Ph.D. study at Monash University, Australia. 
This work was also supported by the Australian Research Council (ARC) under a Laureate Fellowship project FL190100035, a Discovery Early Career Researcher Award (DECRA) project DE200100016, and a Discovery project DP200100020.

\small
 \appendices
\section{Initial data extracted for`London Restaurants Bars \& Pubs +' app in REACT`}
\label{Appedix-A-Chabada}

\textbf{\flushleft Method Name:}  create Parcel create Parcel Array create Parcel Array close flush write describe Contents write To Parcel describe Contents write To Parcel create Parcel Array describe Contents write To Parcel dump Activity Result Back Pressed Configuration Changed Create Create Panel Create Destroy Key Low Memory Selected Panel Closed Pause Post Resume Prepare Panel Resume Retain Non Configuration Instance Save Instance State Start Stop start Activity For Result create Parcel Array Animation End Animation Repeat Animation Start dispatch Restore Instance State dispatch Save Instance State create Parcel Array create Parcel Array describe Contents write To Parcel equals Code Configuration Changed Create Context Low Memory Interpolation Scroll State Scrolling Cache Enabled add Focusables add Touchables add check Params compute Scroll dispatch Key Event dispatch Populate Accessibility Event draw drawable State Changed generate Default Params generate Params generate Params Adapter Current Offscreen Page Limit Page Margin Attached To Winw Draw Intercept Touch Event Measure Request Focus In Descendants Restore Instance State Save Instance State Size Changed Touch Event Adapter Current Offscreen Page Limit On Adapter Change Listener On Page Change Listener Page Margin Page Margin Page Margin very Changed Invalidated compare write To Parcel varargs In Background Cancelled Pre Execute varargs In Background Cancelled Pre Execute accept Post Execute uncaught Exception Bind Destroy Low Memory Start Unbind process Content Content Length Page Finished should Override Url Loading Configuration Changed Create Key Pause Receive Location Changed Provider Disabled Provider Enabled Status Changed should Override Url Loading process varargs In Background Post Execute varargs In Background Post Execute Configuration Changed Create Key User Leave Scale Scale Begin Scale End Create Measure Create Upgrade varargs varargs In Background Of varargs varargs In Background draw Compass draw My Location Location Changed Of bulk Insert delete Type insert Create query update Checked Changed Checked Changed Create Count Id is Enabled is Route Displayed Back Pressed Create Create Dialog Key New Intent Pause Prepare Dialog Resume Save Instance State Search Requested User Leave Post Execute Pre Execute varargs varargs In Background Post Execute Pre Execute Count Id Of Count Id Type Type Count is Enabled Fling Long Press Scroll Show Press Single Tap Up Touch Event Create Create Cancel draw Of available close mark mark Supported read read read re skip Double Tap Double Tap Event Single Tap Confirmed Cancel Create Dismiss Create draw Tap Post Execute Pre Execute Of Create 

\textbf{\\GUI Text:} ii Th ia ii mM AN Bg ae iil Th Be fi OW  fil sa be bad Mi Fabs fil lis
\textbf{\\\\\\XML Data Value:}  Settings Find your location Name Elevation Distance Address Duration Layers Satellite Traffic Settings Current location not yet available Please wait Nearest POIs Not available Please wait Computing distances NA Im in car Im on bike Im walking New proximity alert created Proximity alert removed Could not create proximity alert Map POIs Screen interaction Keep screen on Distance calculation Fast Accurate OK Cancel Download failed Please check your internet connection Google Navigation is not installed Google Streetview is not installed Streetview Navigation Route Route distance is too big Loading data Please wait This application requires working data connection Exit Network failure Navigation mode These options are not available in the current version Tap here to install this app Google Play Store is not installed You might need There are no recommended apps for you at this moment Tap to install in order to view enhanced results To view more results tap here to install this app More options To have full layers access tap here to install this app Distance units Metric Imperial Others Art Galleries Beaches Bridges Buildings Casinos Cinemas Convention Centers Educational Sites Fountains Game Centers Gardens Health Clubs Historic Sites Landmarks Libraries Monuments Museums Neighborhoods Parks Performances Religious Sites SPA Sports Theaters Theme Parks Tours Tourist Centers Transportation Waterfalls Others Star Stars Stars Stars Stars Stars Stars Stars Stars Others Bed amp Breakfast Lodging Apartments Houses Rooms Others Afghan African Albanian American Argentinian Armenian Asian Australian Austrian Bar Barbecue Belgian Bistro Brazilian British Burmese Cambodian Canadian Caribbean Chilean Chinese Coffee Colombian Creperie Croatian Cuban Czech Dessert Dinner Dutch Ecuadorian Egyptian English Ethiopian European Fast Food French German Greek Hungarian Ice Cream Indian Indonesian International Irish Israeli Italian Jamaican Japanese Korean Lebanese Malaysian Mediterranean Mexican Mongolian Moroccan Nepali Organic Others Pakistani Pasta Persian Philippine Pizza Polish Portuguese Pub Romanian Russian Scottish Seafood Spanish Sushi Swedish Swiss Taiwanese Thai Tibetan Tunisian Turkish Vegetarian Vietnamese London RestaurantsampPubs London RestaurantsampPubs

\end{document}